\documentclass{article}
\usepackage{spconf,amsmath,graphicx}
\usepackage{multirow}
\usepackage{lscape,array}
\usepackage{booktabs}
\usepackage{pifont}
\usepackage{ragged2e}
\usepackage{amssymb}
\usepackage{enumitem}
\usepackage{subcaption}
\usepackage{amssymb}
\makeatletter

\newcommand{\Rmnum}[1]{\expandafter\@slowromancap\romannumeral #1@}
\makeatother

\title{Probing Self-supervised Learning Models with 
Target Speech Extraction}
\name{\begin{tabular}{c}Junyi Peng$^{1}$, Marc Delcroix$^{2}$, Tsubasa Ochiai$^{2}$, Old\v{r}ich Plchot$^{1}$, Takanori Ashihara$^{2}$, \\
Shoko Araki$^{2}$, Jan \v{C}ernocký$^{1}$\end{tabular}}

\address{
$^1$Brno University of Technology, Faculty of Information Technology, Speech@FIT, Czechia \\
$^2$NTT Corporation, Japan\\
}
\newcolumntype{L}[1]{>{\raggedright\let\newline\\\arraybackslash\hspace{0pt}}m{#1}}
\newcolumntype{C}[1]{>{\centering\let\newline\\\arraybackslash\hspace{0pt}}m{#1}}
\newcolumntype{R}[1]{>{\raggedleft\let\newline\\\arraybackslash\hspace{0pt}}m{#1}}

\ninept
\begin{document}
\maketitle
\begin{abstract}
Large-scale pre-trained self-supervised learning (SSL) models have shown remarkable advancements in speech-related tasks. However, the utilization of these models in complex multi-talker scenarios, such as extracting a target speaker in a mixture, is yet to be fully evaluated.
In this paper, we introduce target speech extraction (TSE) as a novel downstream task to evaluate the feature extraction capabilities of pre-trained SSL models. TSE uniquely requires both speaker identification and speech separation, distinguishing it from other tasks in the Speech processing Universal PERformance  Benchmark (SUPERB) evaluation. Specifically, we propose a TSE downstream model composed of two lightweight task-oriented modules based on the same frozen SSL model. One module functions as a speaker encoder to obtain target speaker information from an enrollment speech, while the other estimates the target speaker's mask to extract its speech from the mixture. Experimental results on the Libri2mix datasets reveal the relevance of the TSE downstream task to probe SSL models, as its performance cannot be simply deduced from other related tasks such as speaker verification and separation.
\end{abstract}
\section{Introduction}
\label{sec:intro}
Transformer models, empowered by self-supervised learning (SSL) \cite{hsu2021hubert, baevski2020wav2vec, chen2022wavlm, baevski2022data2vec}, have recently marked significant achievements in the field of speech processing, including automatic speech recognition (ASR) \cite{li2023parameter}, speaker verification (SV) \cite{vaessen2022fine, chen2022large, peng2023attention}, and speech enhancement (SE) \cite{song2023exploring, hung2022boosting}. The robustness and generalization abilities of these models are attributed to their capacity to extract general-purpose features through the SSL paradigm on large-scale datasets \cite{peng2023improving}. 

To quantitatively evaluate the SSL models for various speech tasks, benchmarks such as the Speech processing Universal PERformance  Benchmark (SUPERB) and its multilingual variant have been proposed \cite{yang2021superb, tsai2022superb, shi2023ml}. 
In these benchmarks, SSL models are evaluated on several downstream tasks using lightweight task-specific models that rely on input features derived from the layer-wise outputs of the frozen pre-trained SSL models. SUPERB covers diverse downstream tasks, including ASR, SV, SE, speech separation, etc.
%In these benchmarks, a lightweight, task-specific model is integrated with the layer-wise outputs of a frozen pre-trained SSL model during fine-tuning. 

Recently, the problem of Target Speech Extraction (TSE), defined as the process of isolating the speech signal of a target speaker from a multi-talker mixture using auxiliary cues \cite{vzmolikova2023TSEoverview}, has attracted significant interest \cite{ge2020spex, kamo23_interspeech}. This task not only requires speech separation but also the precise identification of the target speaker. Such a dual requirement makes TSE a valuable candidate for evaluating the capabilities of SSL models in extracting fine-grained (acoustic) features and understanding speaker-specific context. However, SUPERB does not include a TSE downstream task. %has not been extensively explored as a downstream task within the SUPERB framework.

%Although these benchmarks, including speech separation tasks, have validated the adaptability of SSL models in a wide range of scenarios, their effectiveness in more challenging conditions, such as multi-talker environments still requires further exploration. Unlike speech separation tasks that aim to disentangle each speaker’s speech from overlapping speech. Multi-talker scenarios present challenges that extend beyond simple separation. This study focuses on target speech extraction (TSE), a process of isolating a speaker of interest from a mixture, guided by auxiliary speech. TSE requires not only separation but also the accurate identification and extraction of a specific speaker's speech . These additional demands set TSE tasks apart from separation tasks, highlighting the necessity of evaluating SSL models for their effectiveness in such challenging multi-talker scenarios.
%Specifically, most SSL models, typically pre-trained on single-source data, have shown remarkable performance in various single-talker tasks but are limited tested in multi-talker mixtures, especially of given auxiliary speech to identify the target speaker . 
%Furthermore, current SUPERB-style downstream tasks suffer from three constraints: (\rmnum{1}) Typically focusing on a single objective; (\rmnum{2}) The majority are based on clean speech, assuming that each recording contains only one speaker; (\rmnum{3}) Neglecting the interconnections between different speech tasks.

In this paper, we introduce a novel TSE downstream task, following SUPERB principles. Specifically, we build an SSL-based \emph{extractor} model that processes a speech mixture to estimate the target speech. This process is conditioned on a target speaker embedding obtained by a \emph{speaker encoder} using the enrollment speech of the target speaker. Both the extractor and speaker encoder are derived from the same pre-trained SSL model. With this new downstream task, we aim to evaluate pre-trained SSL models from a new perspective and answer the following research question: \emph{Is TSE performance governed by the performance of SSL models on the related SV and separation tasks?}

There are a few works that use pre-trained SSL models for TSE \cite{liu2023quantitative, peng2023icassp}.
% Here should mention previous works.
In \cite{liu2023quantitative}, a pre-trained SSL model was explored to extract target speaker embeddings from enrollment speech for TSE, but not for the extraction module. It yielded a marginal improvement over FBANK features. In \cite{peng2023icassp}, an SSL model was employed for encoding both mixtures and speaker enrollment. While it briefly introduced a SUPERB-style downstream model, it predominantly focused on the integration of SSL representations into existing TSE systems (i.e., TD-SpeakerBeam \cite{delcroix2020improving}). Compared to that work, this paper makes the following key contributions:
% With this new benchmark, we conduct a comparative evaluation of the robustness of current SSL models along multiple factors. Firstly, we provide a detailed study of the various fusion techniques, which is used to combine mixture representation and target-speaker-related features. Secondly, we explore the efficiency and effectiveness of SSL-based TSE systems compared to current TSE systems trained from scratch such as TD-Speakerbeam. Finally, we further investigate a fine-tuning strategy to unlock the potential of pre-trained SSL models and improve TSE performance.
%Our key contributions are summarized as follows:
% \begin{itemize}
\begin{itemize}
\item \textbf{SUPERB-TSE System:}  We introduce a novel TSE task for evaluating pre-trained SSL models following principles from SUPERB. With this new task, we investigate various implementation choices for the downstream model, highlighting SSL's potential in extracting the target speaker's speech from mixtures. 

\item \textbf{Comparative Analysis:} We benchmark nine well-known large-scale SSL speech models and three Whisper models \cite{radford2023robust} with our proposed TSE downstream task. 
% Additionally, we explore the impact of downstream data size and various acoustic environments on the TSE performance.

\item \textbf{Performance Correlation:} Our observations reveal that the performance of TSE tasks cannot be simply inferred from the performance on the isolated SV and Separation downstream tasks, suggesting a more intricate relationship between these tasks.

\item \textbf{Comparison with TD-SpeakerBeam:} We compare the performance of the SUPERB-TSE model with a strong TSE system (TD-SpeakerBeam), in terms of training time and performance. It shows that while the SSL-based system enables fast training, there is significant room for further improvement.

%\item Extensive experiments on both Libri2mix and Libri3mix datasets demonstrate the superior performance of our proposed lightweight SSL-based TSE model.

\end{itemize}

\begin{figure}[t]
    \centering
    \includegraphics[width=0.99\linewidth]{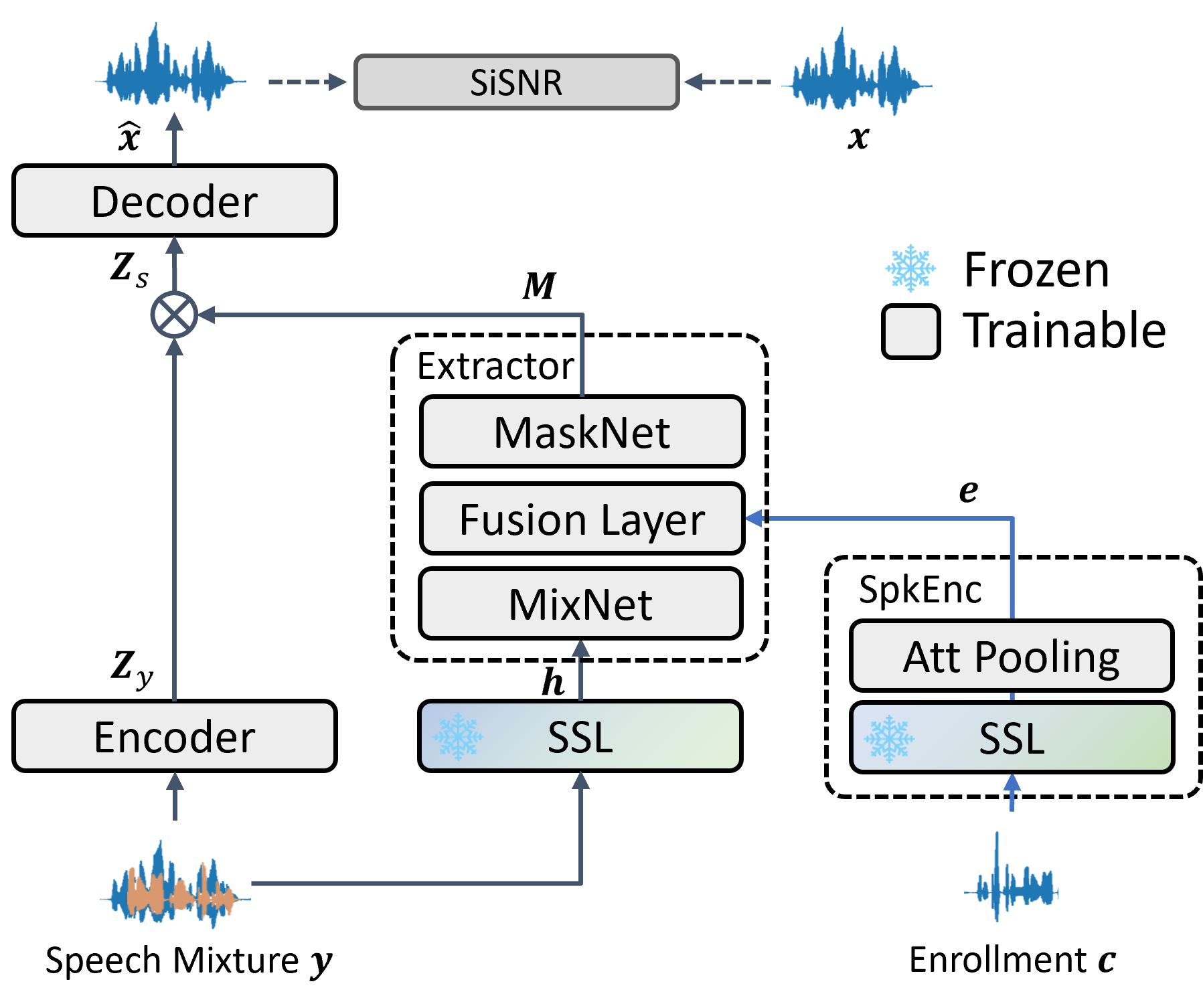}
    \vspace{-0.2cm}
    \caption{Architecture of proposed SSL-based TSE system.}
    \label{fig:sys}
    \vspace{-0.6cm}
\end{figure}

\section{SUPERB-style Target Speech Extraction}
In this section, we introduce the downstream TSE model used to probe SSL models. Figure \ref{fig:sys} shows the architecture of the TSE model, emphasizing the pre-trained SSL models.
The TSE problem consists of extracting the speech of a target speaker, $\mathbf{x}$, in a mixture, $\mathbf{y} = \mathbf{x}+\mathbf{i}$, where $\mathbf{i}$ consists of interference speakers and noise. We rely on an enrollment utterance, $\mathbf{c}$, to identify the target speaker.
%formalize each self-pretraining component in detail, as illustrated in Fig \ref{fig:sys}. 

The proposed SSL-based TSE system consists of four main blocks: the encoder, decoder, speaker encoder (SpkEnc), and extractor, as a typical TSE system \cite{vzmolikova2023TSEoverview}. The encoder transforms the input speech mixture $\mathbf{y}$ into a sequence of features $\mathbf{Z}_y$ by $\mathbf{Z}_y=\text{Encoder}(\mathbf{y})$. SpkEnc is responsible for computing a speaker embedding vector, $\mathbf{e}$, from the enrollment speech, $\mathbf{c}$, as $\mathbf{e}=\text{SpkEnc}(\mathbf{c})$. 
Subsequently, the extractor computes the target speech mask $\mathbf{M}$ within the feature domain $\mathbf{Z}_y$ from SSL features $\mathbf{h}$ and the  target speaker embedding $\mathbf{e}$, as $\mathbf{M}=\text{Extractor}(\mathbf{h},\mathbf{e})$. 
Here, following the standard SUPERB approach, the SSL features consist of a weighted sum of the outputs of the Transformer blocks.
We then obtain a feature representation of the target speech, $\mathbf{Z}_s$, by applying the mask on the features of the mixture, as  $\mathbf{Z}_s=\mathbf{M} \odot \mathbf{Z}_y$, where $\odot$ denotes Hadamard multiplication operation. 
Finally, the decoder converts $\mathbf{Z}_s$ back to the time domain, to obtain the target speech signal as, $\hat{\mathbf{x}}=\text{Decoder}(\mathbf{Z}_s)$.
%Subsequently, the extractor computes the target speech mask $\mathbf{M}$ within the feature domain $\mathbf{Z}_y$, utilizing the layer-wise SSL features $\mathbf{h}$ and guided by  target speaker embedding $\mathbf{e}$, expressed as $\mathbf{M}=\text{Extractor}(\mathbf{h},\mathbf{e})$. 
%Finally, the decoder processes the masked features $\mathbf{Z}_s=\mathbf{M} \odot \mathbf{Z}_y$, where $\odot$ denotes Hadamard multiplication operation, to generate the estimated target speech signal $\hat{\mathbf{x}}=\text{Decoder}(\mathbf{Z}_s)$. 
\subsection{Encoder/Decoder}
The Encoder module operates on the raw waveform by utilizing a set of time-domain finite impulse response filters. We explore two options for the encoder and decoder.

First, we can use Short Time Fourier Transform (STFT) and inverse STFT (iSTFT) for the encoder and decoder, respectively. 
%These filters can be initialized and kept frozen with pre-defined complex-valued exponential functions corresponding to the Short Time Fourier Transform (STFT). In this case, the decoder performs the inverse STFT (iSTFT), converting the target speech spectrogram  $\mathbf{Z}_s$ back into the time domain. 
Alternatively, the filter bank can also be randomly initialized and then jointly optimized with the entire TSE system, allowing the filters to focus on task-related frequency bands \cite{luo2018tasnet}. Accordingly, the Decoder is implemented by using a deconvolution layer that up-samples the target speaker features $\mathbf{Z}_s$ back into the time-domain waveform.

\subsection{SSL-based Speaker Encoder}
In SUPERB-style SV downstream tasks, speaker representation is obtained using a time-delay neural network (TDNN)-based speaker extractor (i.e. X-vector \cite{snyder2018x}) from the weighted sum of layer-wise SSL features. To construct a more lightweight module, we proposed an attentive pooling called multi-head factorized attentive pooling (MHFA) to extract speaker information \cite{peng2023attention}, as shown in Figure \ref{fig:sys}. The SSL-MHFA utilizes two sets of normalized layer-wise weights to generate attention maps and compressed features, respectively. These are designed to encode speaker-discriminative information and phonetic information, respectively. 
We obtain the speaker embedding vector by aggregating the compressed features over frames and then project the resulting vector to a lower-dimensional space using a linear layer.
%Through aggregating over frames and projecting the resulting vector to a lower-dimensional space using a linear layer, the speaker embedding is obtained. 
This method allows each attention head to focus on specific phonetic units, resulting in a speaker embedding that is robust to phonetic variability.

In the proposed downstream TSE model, we use the SSL-MHFA module to compute the target speaker embedding vector, $\mathbf{e}$. Note that this module is trained jointly with the other components of the TSE model without any speaker identification loss.

\begin{table}[]
% \vspace{-0.6cm}
\caption{Different fusion methods in TSE. FiLM: Feature-wise linear modulation, which uses two vectors $\mathbf{e}_1$ and $\mathbf{e}_2$ obtained by projecting the embedding vector, $\mathbf{e}$, with two learnable linear layers.
%we double the final embedding dimension and decompose it into $\mathbf{e}_1$ and $\mathbf{e}_2$.
}
\vspace{-0.4cm}
\label{tab:fusion_layers}
\centering
\begin{tabular}{llc}
\toprule
Fusion Method & Implementation & SI-SDRi$\uparrow$ \\ \midrule
Addition & $\mathbf{Z}_f=\mathbf{Z}_{mix} + \mathbf{e}$ &9.13    \\
Multiplication & $\mathbf{Z}_f=\mathbf{Z}_{mix} \odot \mathbf{e}$ & \textbf{9.96}   \\
Concatenation & $\mathbf{Z}_f=concat(\mathbf{Z}_{mix}, \mathbf{e})$ & 8.88  \\
FiLM & $\mathbf{Z}_f=\mathbf{Z}_{mix}\odot\mathbf{e}_1 + \mathbf{e}_2$& 8.16  \\
\bottomrule
\end{tabular}
\vspace{-0.3cm}
\end{table}

\begin{table*}[t]
\caption{Evaluating different TSE model configurations in Libri2mix (16kHz-min). For a fair comparison, we use the layer-wise outputs of the pre-trained WavLM Base Plus model as SSL features. }
\vspace{-0.4cm}
\label{tab:my-table2}
\centering
%\begin{tabular}{cccccrrr}
\begin{tabular}{llllllccc}
\toprule
System&Encoder/Decoder & Extractor & SpkEnc & Mask Type      & Objective & SI-SDRi$\uparrow$ & STOI(\%)$\uparrow$ & PESQ$\uparrow$ \\ 
\midrule
1&STFT/iSTFT      & STFT      & STFT   & Magnitude Mask & MSE       &5.96         &79.55      &1.42      \\
2&STFT/iSTFT      & STFT      & SSL    & Magnitude Mask & MSE       &7.42         &81.75      &1.51      \\
3&STFT/iSTFT      & SSL       & STFT    & Magnitude Mask & MSE      &8.70         &85.03      &1.83      \\
4&STFT/iSTFT      & SSL       & SSL    & Magnitude Mask & MSE       &9.96         &87.79      &\textbf{1.97}      \\
\midrule
5&STFT/iSTFT      & SSL       & SSL    & Magnitude Mask & SI-SDR    &10.66         &88.74      &1.91     \\
6&STFT/iSTFT      & SSL       & SSL    & Complex Mask   & SI-SDR    &10.61         &88.84      &1.91      \\
\midrule
7&Conv1D/DeConv1D & SSL       & SSL    & Encoder-domain Mask           & SI-SDR    &\textbf{11.04}         &\textbf{89.47}      &1.93 \\
\bottomrule
\end{tabular}
\end{table*}

\begin{table*}[t]
\caption{Comparison of different general-purpose speech models for TSE, SV, and separation downstream tasks. For Whisper models, we only use the audio encoder. In the fine-tuning case, initializing from a converged model with a frozen SSL (e.g. WavLM Base Plus), we unfreeze the SSL and further train the entire system for 20 epochs.}
\vspace{-0.3cm}
\label{tab:ssl_comparison}
\centering
% \scalebox{0.8}{
\begin{tabular}{llccccccc}
\toprule
\multirow{2}{*}{Upstream} & \multirow{2}{*}{\#Params} & \multicolumn{4}{c}{TSE} & \multicolumn{1}{c}{SV (MHFA)} & \multicolumn{1}{c}{SV (Xvector)} & \multicolumn{1}{c}{Separation} \\ 
\cmidrule(lr){3-6} \cmidrule(lr){7-7} \cmidrule(lr){8-8} \cmidrule(lr){9-9} 
                          & & SI-SDRi$\uparrow$  & STOI (\%)$\uparrow$  & PESQ$\uparrow$ & FR(\%)$\downarrow$  & EER(\%) $\downarrow$ & EER(\%) $\downarrow$  & SI-SDRi$\uparrow$               \\
\midrule
Whisper-Base & 20.59M & 9.25 & 86.13 & 1.71 & 6.11 & 3.39 & 9.55 & 9.76 \\
Whisper-Small & 88.15M & \underline{10.28} & \underline{88.79}& 1.82 & \underline{4.45} & \underline{2.55} & 9.19 & \underline{11.06} \\
Whisper-Medium & 307.22M & 9.94& 87.26 & \underline{1.85} & 6.50 & 4.22 & \underline{8.66} & 10.94\\
\midrule
Data2vec Base     &93.84M        &9.43      &\underline{86.21}       &1.72      &\underline{5.45}      &3.51   &\underline{6.79}                      &9.95                        \\
Data2vec Large   &314.30M         &\underline{9.55}      &86.11       &\underline{1.77}      &7.46      &\underline{2.59}    &7.61                    &\underline{10.81 }                      \\
\midrule
wav2vec 2.0 Base  &95.00M        &\underline{9.52}      &\underline{86.34}       &1.72      &\underline{5.23}      &3.53  &\underline{6.10}                      &10.01                        \\
wav2vec 2.0 Large  &317.38M       &8.40      &84.24       &\underline{1.74}      &9.31      &\underline{3.04}     &6.38&\underline{10.31}                        \\
\midrule
Hubert Base      &94.68M         &\underline{9.62}      &\underline{86.69}       &1.74      &\underline{4.56}      &3.06   &\underline{5.30}                      &10.01                        \\
Hubert Large     &316.61M          &9.03      &85.41       &\underline{1.88}      &8.73      &\underline{2.94}   &5.82&\underline{10.95}                        \\
\midrule
WavLM Base       &94.70M         &10.03      &87.99      &1.84      &3.71      &2.71  &5.36                      &10.80                        \\
WavLM Base Plus   &94.70M        &\textbf{11.04}         &\textbf{89.47}       &1.93 &\textbf{3.45}       &\textbf{2.03}   &\textbf{4.39}                     &11.41                        \\
WavLM Large     &316.62M          &9.73          &86.53       &\textbf{2.04} &7.96      &2.30   &4.87&\textbf{11.87}                        \\
\midrule
\multicolumn{2}{l}{WavLM Base Plus [Fine-tuning]} & 11.51 & 90.08 & 2.01 & 3.23 & - & -& -\\ 
\multicolumn{2}{l}{TD-SpeakerBeam} & 13.03 & 90.63 & 2.21 & 4.85 & - & -& -\\
\bottomrule
\end{tabular}
% }
\vspace{-0.5cm}
\end{table*}

\subsection{SSL-based Extractor}
The extractor is composed of three sub-modules: a mixture encoder (MixNet), a fusion layer, and a target mask generator (MaskNet), as illustrated in Figure \ref{fig:sys}. It accepts the SSL features, $\mathbf{h}$, as input. The processing is conditioned on the target speaker through a fusion layer that combines the speaker embedding $\mathbf{e}$ and the output of the MixNet, $\mathbf{Z}_{mix}$, as $\mathbf{Z}_f=\text{Fusion}(\mathbf{Z}_{mix}, \mathbf{e})$.  
Finally, MaskNet computes the mask, $\mathbf{M}$ from $\mathbf{Z}_f$. 

In this study, MixNet is implemented by a single BLSTM layer, while Masknet employs two BLSTM layers. We explore various options for the fusion layer \cite{vzmolikova2019speakerbeam}, summarized in Table \ref{tab:fusion_layers}.
%We then multiply the speech mixture features $\mathbf{Z}_y$ with the estimated mask $\mathbf{M}$ and feed it to the decoder to produce the waveform of the extracted speech $\hat{x}$. 

\section{Experiments}
% \vspace{-2mm}
We perform three sets of experiments. First, we investigate the configuration for the TSE downstream model. We then probe various pre-trained SSL models on the proposed TSE downstream task. Finally, we compare the performance with a powerful TSE system.
\subsection{Experiment Setup}
% \vspace{-1.5mm}
\textbf{Datasets:}
%\subsubsection{Datasets}
In this work, we conduct comparative experiments across three downstream tasks: TSE, SV, and Separation. For TSE, we use Libri2Mix \cite{cosentino2020librimix}, consisting of simulated mixtures of two speakers. Following the enrollment speech preparation in TD-SpeakerBeam\footnote{https://github.com/BUTSpeechFIT/speakerbeam} with 16kHz sampling rate, the dataset is partitioned into three subsets: train-100, valid, and test. 
Our proposed TSE downstream task relies on Libri2mix with train-100 for faster experimental turnover. Consequently, all experiments in the paper rely on this configuration, except when specifically specified in the exploration experiments in Section \ref{sec:exploration}. % allAll probing experiments use Libri2mix with train-100 for the probing experiments, and only use Libri3mix and 
Regarding SV, all models are trained using the VoxCeleb1 dataset and evaluated on VoxCeleb1-O \cite{nagrani2017voxceleb}. The speakers in the training and evaluation sets are different.
%There are no overlapped speakers between training and evaluation. 
For Separation, we evaluate the performance of SSL models on Libri2mix.

\noindent\textbf{Implementation details:} 
%\subsubsection{Configuration}
In the system using STFT/iSTFT for Encoder/Decoder, the window size and the number of FFT points are set to 1024 with a stride of 320. This stride aligns with the downsampling rate in the SSL model. Additionally, the dimension of BLSTM is 512. It is noted that this configuration is coherent with the SUPERB's Separation downstream task hyper-parameters \cite{tsai2022superb}. For the systems with learnable kernels, using Conv1D and DeConv1D, the kernel size is set to 1024 with a stride of 320, and the number of filters is 512. 

%For the speaker encoder, when processing the spectrogram of enrollment speech, a three-layer BLSTM followed by average pooling is used to obtain speaker embedding. 
For the speaker encoder, when the input consists of STFT, we process the magnitude of the STFT coefficients with a three-layer BLSTM followed by average pooling to derive the speaker embedding vectors. 
Conversely, for SSL features, we employ MHFA for speaker embedding extraction. The MHFA is configured with four heads and a compression layer with a dimension of 128. 

We trained all modules of the TSE model jointly using the Mean Squared Error (MSE) or scale-invariant signal-to-distortion ratio (SI-SDR) loss between the reference and estimated target speech \cite{kolbaek2020loss}. Note that the MSE is computed in the spectral domain, while SI-SDR is computed in the time domain. We used the Adam optimizer and trained the model for 200 epochs.

For the SV task, we use AM-softmax loss with a scale of 30 and a margin of 0.4 as the objective function. We set the number of heads to 32 for MHFA, resulting in a model size of 2.23M parameters compared to the 5.71M parameters of Xvector.

For the separation experiments, we simply employ the default configuration in SUPERB, which consists of a three-layer BLSTM.

\noindent\textbf{Performance Metrics:} 
%\subsubsection{Performance Metrics:}
For TSE, we measure performance in terms of scale-invariant signal-to-distortion ratio (SI-SDR) improvement (SI-SDRi), perceptual evaluation of speech quality (PESQ), short-time objective intelligibility (STOI), and Failure rate (FR) \cite{delcroix2022listen}. FR measures the proportion of test samples with an SI-SDRi below 1 dB. Failures typically occur when the TSE system extracts the incorrect speaker or outputs the mixture. For SV, we calculate the Equal Error Rate (EER).

%\subsection{Comparison of different fusion strategies}
\subsection{Analysis of downstream model configuration}
\label{sec:analysis}
We first perform several experiments to find an effective configuration for the TSE downstream model. Here, we use WavLM Base Plus
%\footnote{https://github.com/microsoft/unilm/tree/master/wavlm} 
as the upstream model and  STFT/iSTFT as the encoder/decoder unless specified.

In Table \ref{tab:fusion_layers}, we evaluate the effectiveness of various fusion strategies employed in TSE. We observe that the multiplication strategy outperforms other approaches. Consequently, we use it in all subsequent experiments.
%The \textit{Multiplication} strategy outperforms the \textit{Addition} approach, achieving the highest SI-SDRi of 9.96 dB, indicating a more effective integration of speaker characteristics within the speech mixture. Conversely, the \textit{Concatenation} strategy results in a slight decrease in performance across all metrics compared to \textit{Addition}. Notably, \textit{FiLM}, which has theoretical advantages, is the worst with the lowest scores in all evaluated metrics, which may suggest the linear modulation strategy of \textit{FiLM} is insufficient to disentangle the target speech effectively.

%\subsection{Performance Analysis of TSE Configurations}
In Table \ref{tab:my-table2}, we present an extensive evaluation of different TSE model configurations on the Libri2mix dataset.
First, we compare using SSL features with STFT coefficients for the extractor and SpkEnc (systems 1 to 4). We observe that using SSL features for both the extractor and SpkEnc (system 4) results in significantly higher extraction performance. %This confirms that SSL features contain more comprehensive information beneficial for speech-related downstream tasks compared to hand-crafted features such as STFT coefficients. 
Next, we explore the use of the time domain SI-SDR loss and using magnitude or complex masks (systems 5 and 6). We found that using SI-SDR loss significantly improves SI-SDRi and STOI scores. However, using complex masks (system 6) had little impact on the results. 
Finally, we replaced the STFT/iSTFT encoder/decoder modules with learnable ones (system 7). This again improved SI-SDR and STOI scores. We use that configuration in subsequent experiments. %This suggests learnable kernels might be able to capture frequency bands or temporal patterns that are relevant for TSE tasks from mixture waveform.

\begin{figure}[t]
    \centering
    \begin{subfigure}[b]{0.95\linewidth}
    \includegraphics[width=0.95\linewidth]{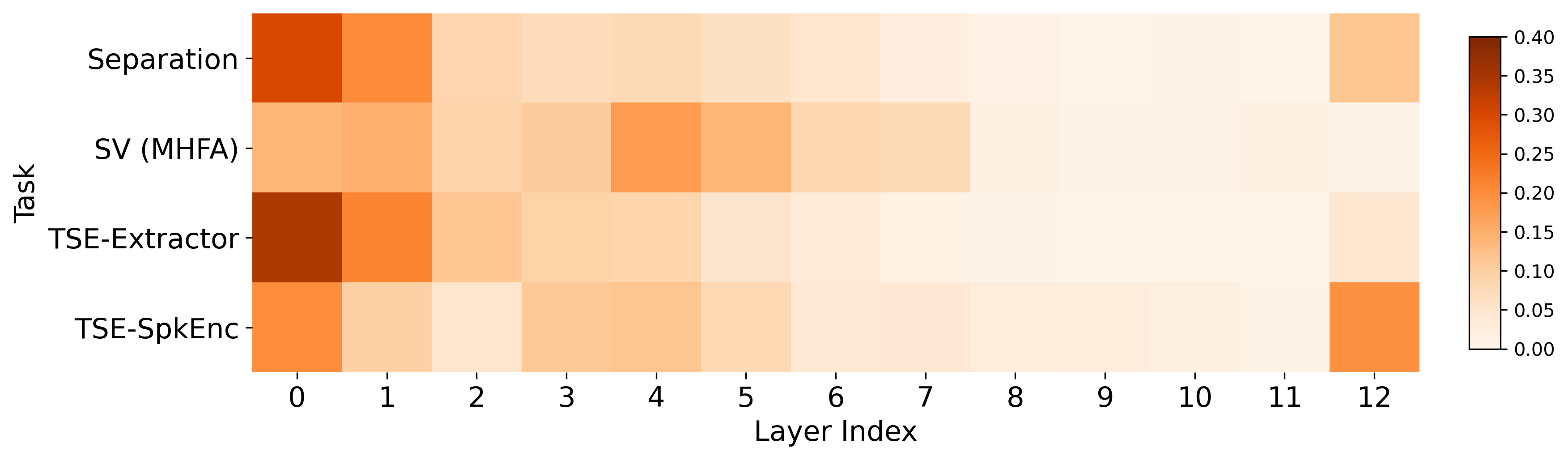}
    \caption{WavLM Base Plus}
    \end{subfigure}
    \begin{subfigure}[b]{0.95\linewidth}
    \includegraphics[width=0.95\linewidth]{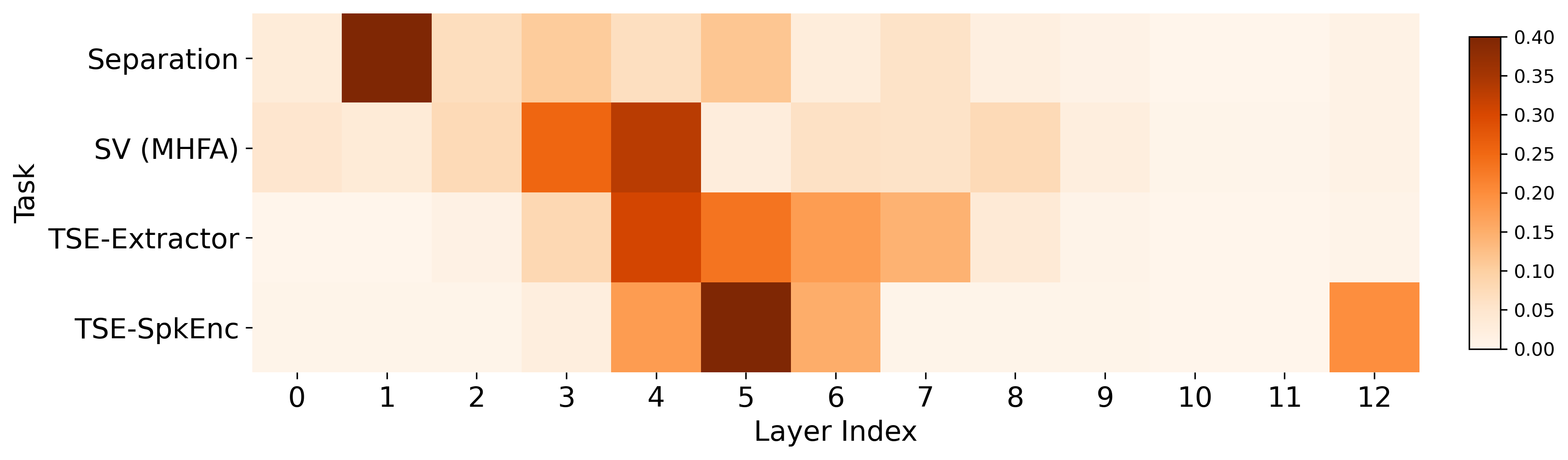}
    \caption{Whisper Small}
    \end{subfigure}
    \caption{The weight distribution of Transformer layers. Note that $0$-th Transformer layer denotes the output of the CNN encoder, which is also the input of the $1$-st Transformer layer. }
    \label{fig:weights}
    \vspace{-0.4cm}
\end{figure}

\begin{figure}[t]
    \centering
    \includegraphics[width=0.87\linewidth]{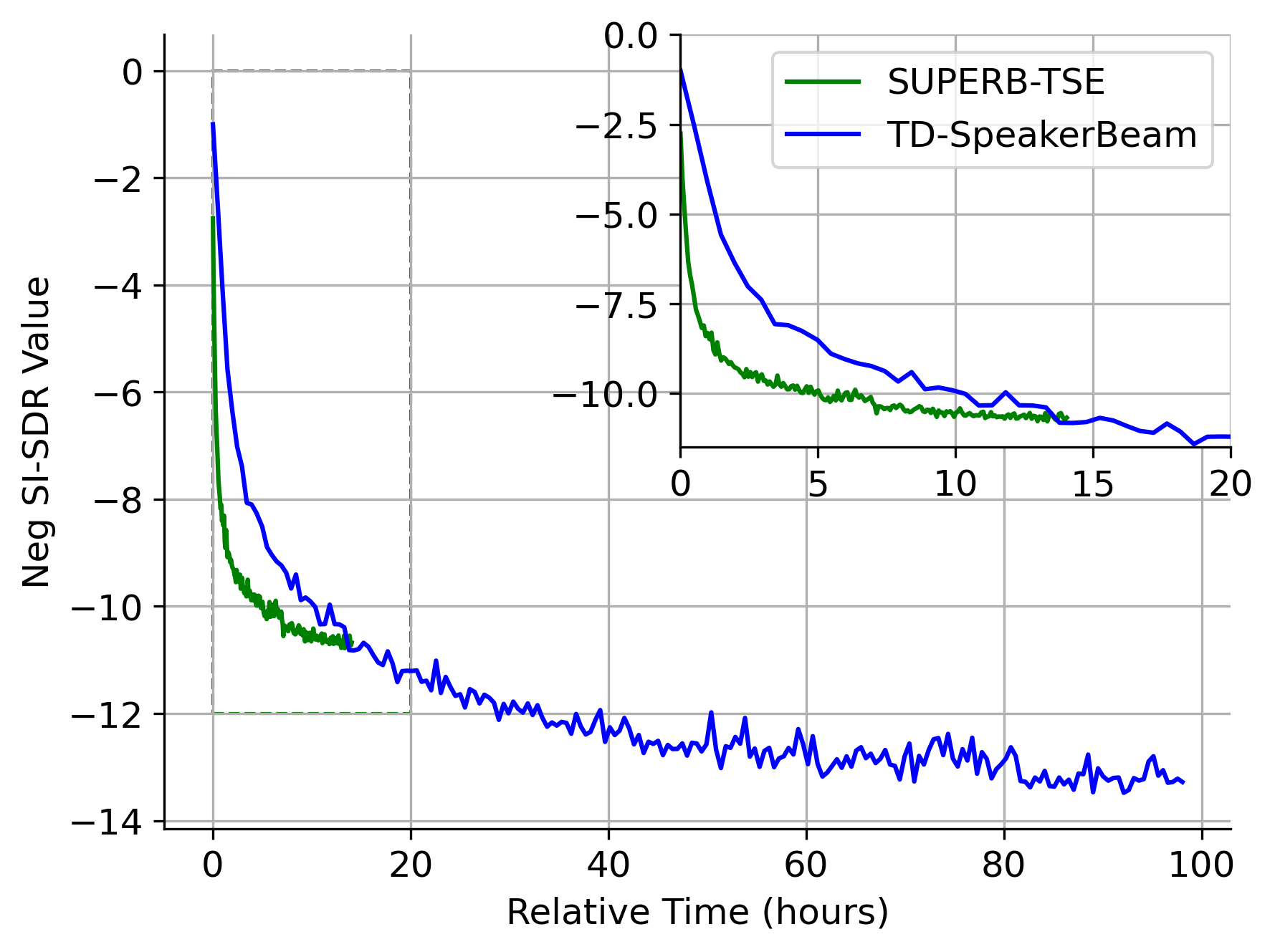}
        % \vspace{-0.4cm}
    \caption{Comparison of SUPERB-TSE and TD-SpeakerBeam training curves, i.e.,  the validation loss (negative SI-SDR) as a function of the training time (time $\times$ \# GPUs). %The x-axis represents the number of GPUs multiplied by the training time. The y-axis indicates the loss on validation (negative SI-SDR), where lower values mean better performance.
}
    \label{fig:enter-label1}
    \vspace{-0.4cm}
\end{figure}

%\begin{figure}[t]
%     \centering
%     \includegraphics[width=0.99\linewidth]{Fig/Fig1.png}
%     \caption{Comparative performance of TSE systems across different training datasets. “SUPERB-TSE [FT]” enables fine-tuning of the SSL model (i.e. WavLM Base Plus) with the entire system. 
% }
%     \label{fig:enter-label2}
% \end{figure}

\subsection{Performance Comparison of various SSL models}
In Table \ref{tab:ssl_comparison}, we evaluate the performance of various SSL models (and Whisper encoder) across three tasks: TSE, SV, and speech separation. 
The models are trained with LibriSpeech data and LibriLight \cite{kahn2020libri}. Note that the difference between WavLM Base and Base Plus is the amount of training data, i.e., WavLM Base is trained only with LibriSpeech 960 hours of training data, while WavLM Base Plus is trained using the same data as WavLM Large, i.e., 94k hours.

First, let us look at the SV and separation performance. For SV, we observe that MHFA outperforms Xvector with fewer parameters (Params: 2.31M v.s. 5.71M), suggesting its effectiveness. For SV with MHFA, Large SSL models tend to perform best, except for WavLM where WavLM Base Plus achieves significantly better performance. For separation, WavLM Large outperforms other models, and large SSL models constantly perform better than base ones.

Intuitively, we would expect that SSL models achieving high scores in terms of separation and extraction should be good for TSE. However, we observe that there is a more intricate relation, e.g., Base models usually perform better than Large models for TSE although they perform worse in terms of SV and separation. Compared to other SSL models, WavLM Base Plus achieves the best performance for SV and TSE tasks. This suggests the importance of data augmentation in the pre-training stage to capture robust speech and speaker representations.

These experiments demonstrate that the performance of TSE models is not directly correlated with their performance in Separation and SV tasks. Moreover, as illustrated in Figure \ref{fig:weights},  the distributions for Separation and Extractor, as well as SV and SpkEnc, are distinctly different. However, the two TSE sub-modules show a similar pattern, which might be a result of their joint optimization. This observation further emphasizes the uniqueness of TSE and the necessity for task-specific evaluation of SSL models rather than assuming a universally effective model across various speech tasks.
%In SV, MHFA outperforms Xvector with fewer parameters (Params: 2.31M v.s. 5.71M), suggesting its effectiveness. A Similar trend is observed between Large and Base SSL models in SS and SV tasks, where the model size and pre-trained dataset significantly impact downstream task performance. Compared to other SSL models, WavLM-based systems achieve advanced performance across three downstream tasks, suggesting data augmentation methods in the pre-training stage can improve the robustness of the model. Notably, the effectiveness of SSL models in TSE is not directly correlated with their performance in SS and SV tasks. This emphasizes the uniqueness of TSE and the necessity for task-specific evaluation of SSL models, rather than assuming a universally effective model across various speech tasks.

%\subsection{Exploration study across training process}
\subsection{Comparison with powerful TSE system}
\label{sec:exploration}
Finally, we compare the WavLM-based TSE system (hereafter called SUPERB-TSE), with a strong TSE model, TD-SpeakerBeam \cite{delcroix2020improving}. First, Figure \ref{fig:enter-label1} compares the training speed of these two models.
The SUPERB-TSE system demonstrates faster convergence, achieving a 10 dB SI-SDR within 2-3 hours of training time, in contrast to TD-SpeakerBeam, which requires over 10 hours to achieve similar performance. Moreover, TD-SpeakerBeam reaches full convergence in 100 hours, whereas SUPERB-TSE achieves this in only 14 hours. Moreover, as seen in Table \ref{tab:ssl_comparison}, the failure rate is lower with the WavLM-based TSE system (FR: 3.45 vs 4.85 \%)), which may indicate better speaker identification capabilities. However, despite these benefits, there remains a significant performance gap between SUPERB-TSE and TD-SpeakerBeam systems (SI-SDRi: 11.04 dB v.s. 13.03 dB as shown in Table \ref{tab:ssl_comparison}). %Despite these advancements in training efficiency, there remains a significant performance gap between SUPERB-TSE and TD-SpeakerBeam systems (SI-SDRi: 11.04 dB v.s. 13.03 dB as shown in Table \ref{tab:ssl_comparison}). 
We attribute part of this gap to the very simple model architecture of the downstream model, as well as to the low time resolution of the SSL model, which may not be optimal for speech enhancement \cite{peng2023icassp}.

\section{Conclusions}
In this work, we introduce a novel SSL-based TSE aligned with SUPERB principles. Our comprehensive experiments on Libri2Mix datasets demonstrate that TSE performance cannot be directly inferred from SV and Separation tasks, justifying the importance of including the TSE downstream task when probing SSL models. 
We showed that with careful implementation choices, we can build a relatively strong TSE downstream model, which achieves fast convergence. However, such a simple downstream model still lags behind more powerful TSE systems trained from scratch, such as TD-SpeakerBeam. However, fine-tuning the SSL model for TSE as well as enhancing the temporal resolution of SSL models, may constitute promising research directions to boost TSE performance \cite{peng2023icassp}. 

% We also discussed that the proposed SSL-based TSE system can achieve faster convergence than more complex powerful TD-SpeakerBeam model, the performance still lags behind current TD-SpeakerBeam system trained from scratch.
%Additionally, further fine-tuning SSL models have shown great potential in processing multi-talker scenarios, while also indicating that enhancing the temporal resolution of the model could be a promising direction for future research.

% Below is an example of how to insert images. Delete the ``\vspace'' line,
% uncomment the preceding line ``\centerline...'' and replace ``imageX.ps''
% with a suitable PostScript file name.
% -------------------------------------------------------------------------

%\vfill
%\pagebreak

\vfill\pagebreak

\footnotesize
\bibliographystyle{IEEEbib}
\bibliography{refs}

\begin{thebibliography}{10}

\bibitem{hsu2021hubert}
Wei-Ning Hsu, Benjamin Bolte, Yao-Hung~Hubert Tsai, Kushal Lakhotia, Ruslan Salakhutdinov, and Abdelrahman Mohamed,
\newblock ``Hubert: Self-supervised speech representation learning by masked prediction of hidden units,''
\newblock {\em IEEE/ACM Transactions on Audio, Speech, and Language Processing}, vol. 29, pp. 3451--3460, 2021.

\bibitem{baevski2020wav2vec}
Alexei Baevski, Yuhao Zhou, Abdelrahman Mohamed, and Michael Auli,
\newblock ``wav2vec 2.0: A framework for self-supervised learning of speech representations,''
\newblock {\em Advances in neural information processing systems}, vol. 33, pp. 12449--12460, 2020.

\bibitem{chen2022wavlm}
Sanyuan Chen, Chengyi Wang, Zhengyang Chen, Yu~Wu, Shujie Liu, Zhuo Chen, Jinyu Li, Naoyuki Kanda, Takuya Yoshioka, Xiong Xiao, et~al.,
\newblock ``{WavLM: Large-Scale Self-Supervised Pre-Training for Full Stack Speech Processing},''
\newblock {\em IEEE Journal of Selected Topics in Signal Processing}, vol. 16, no. 6, pp. 1505--1518, 2022.

\bibitem{baevski2022data2vec}
Alexei Baevski, Wei-Ning Hsu, Qiantong Xu, Arun Babu, Jiatao Gu, and Michael Auli,
\newblock ``Data2vec: A general framework for self-supervised learning in speech, vision and language,''
\newblock in {\em International Conference on Machine Learning}. PMLR, 2022, pp. 1298--1312.

\bibitem{li2023parameter}
Zhengyang Li, Thomas Graave, Jing Liu, Timo Lohrenz, Siegfried Kunzmann, and Tim Fingscheidt,
\newblock ``Parameter-efficient cross-language transfer learning for a language-modular audiovisual speech recognition,''
\newblock in {\em 2023 IEEE Automatic Speech Recognition and Understanding Workshop (ASRU)}. IEEE, 2023, pp. 1--8.

\bibitem{vaessen2022fine}
Nik Vaessen and David~A Van~Leeuwen,
\newblock ``Fine-tuning wav2vec2 for speaker recognition,''
\newblock in {\em ICASSP 2022-2022 IEEE International Conference on Acoustics, Speech and Signal Processing (ICASSP)}. IEEE, 2022, pp. 7967--7971.

\bibitem{chen2022large}
Zhengyang Chen, Sanyuan Chen, Yu~Wu, Yao Qian, Chengyi Wang, Shujie Liu, Yanmin Qian, and Michael Zeng,
\newblock ``Large-scale self-supervised speech representation learning for automatic speaker verification,''
\newblock in {\em ICASSP 2022-2022 IEEE International Conference on Acoustics, Speech and Signal Processing (ICASSP)}. IEEE, 2022, pp. 6147--6151.

\bibitem{peng2023attention}
Junyi Peng, Old{\v{r}}ich Plchot, Themos Stafylakis, Ladislav Mo{\v{s}}ner, Luk{\'a}{\v{s}} Burget, and Jan {\v{C}}ernock{\`y},
\newblock ``An attention-based backend allowing efficient fine-tuning of transformer models for speaker verification,''
\newblock in {\em 2022 IEEE Spoken Language Technology Workshop (SLT)}. IEEE, 2023, pp. 555--562.

\bibitem{song2023exploring}
Hyungchan Song, Sanyuan Chen, Zhuo Chen, Yu~Wu, Takuya Yoshioka, Min Tang, Jong~Won Shin, and Shujie Liu,
\newblock ``{Exploring WavLM on Speech Enhancement},''
\newblock in {\em 2022 IEEE Spoken Language Technology Workshop (SLT)}. IEEE, 2023, pp. 451--457.

\bibitem{hung2022boosting}
Kuo-Hsuan Hung, Szu wei Fu, Huan-Hsin Tseng, Hsin-Tien Chiang, Yu~Tsao, and Chii-Wann Lin,
\newblock ``{Boosting Self-Supervised Embeddings for Speech Enhancement},''
\newblock in {\em Proc. Interspeech 2022}, 2022, pp. 186--190.

\bibitem{peng2023improving}
Junyi Peng, Oldřich Plchot, Themos Stafylakis, Ladislav Mosner, Lukáš Burget, and Jan~"Honza" Černocký,
\newblock ``{Improving Speaker Verification with Self-Pretrained Transformer Models},''
\newblock in {\em Proc. INTERSPEECH 2023}, 2023, pp. 5361--5365.

\bibitem{yang2021superb}
Shu wen Yang, Po-Han Chi, Yung-Sung Chuang, Cheng-I~Jeff Lai, Kushal Lakhotia, Yist~Y. Lin, Andy~T. Liu, Jiatong Shi, Xuankai Chang, Guan-Ting Lin, Tzu-Hsien Huang, Wei-Cheng Tseng, {Ko-tik} Lee, Da-Rong Liu, Zili Huang, Shuyan Dong, Shang-Wen Li, Shinji Watanabe, Abdelrahman Mohamed, and {Hung-yi} Lee,
\newblock ``{SUPERB: Speech Processing Universal PERformance Benchmark},''
\newblock in {\em Proc. Interspeech 2021}, 2021, pp. 1194--1198.

\bibitem{tsai2022superb}
Hsiang-Sheng Tsai, Heng-Jui Chang, Wen-Chin Huang, Zili Huang, Kushal Lakhotia, Shu-wen Yang, Shuyan Dong, Andy Liu, Cheng-I Lai, Jiatong Shi, et~al.,
\newblock ``{SUPERB-SG: Enhanced Speech processing Universal PERformance Benchmark for Semantic and Generative Capabilities},''
\newblock in {\em Proceedings of the 60th Annual Meeting of the Association for Computational Linguistics (Volume 1: Long Papers)}, 2022, pp. 8479--8492.

\bibitem{shi2023ml}
Jiatong Shi, Dan Berrebbi, William Chen, Ho-Lam Chung, En-Pei Hu, Wei~Ping Huang, Xuankai Chang, Shang-Wen Li, Abdelrahman Mohamed, Hung-yi Lee, et~al.,
\newblock ``Ml-superb: Multilingual speech universal performance benchmark,''
\newblock {\em arXiv preprint arXiv:2305.10615}, 2023.

\bibitem{vzmolikova2023TSEoverview}
Katerina Zmolikova, Marc Delcroix, Tsubasa Ochiai, Keisuke Kinoshita, Jan Černocký, and Dong Yu,
\newblock ``Neural target speech extraction: An overview,''
\newblock {\em IEEE Signal Processing Magazine}, vol. 40, no. 3, pp. 8--29, 2023.

\bibitem{ge2020spex}
Meng Ge, Chenglin Xu, Longbiao Wang, Eng~Siong Chng, Jianwu Dang, and Haizhou Li,
\newblock ``{SpEx+: A Complete Time Domain Speaker Extraction Network},''
\newblock in {\em Proc. Interspeech 2020}, 2020, pp. 1406--1410.

\bibitem{kamo23_interspeech}
Naoyuki Kamo, Marc Delcroix, and Tomohiro Nakatani,
\newblock ``{Target Speech Extraction with Conditional Diffusion Model},''
\newblock in {\em Proc. INTERSPEECH 2023}, 2023, pp. 176--180.

\bibitem{liu2023quantitative}
Xiaoyu Liu, Xu~Li, and Joan Serr{\`a},
\newblock ``Quantitative evidence on overlooked aspects of enrollment speaker embeddings for target speaker separation,''
\newblock in {\em ICASSP 2023-2023 IEEE International Conference on Acoustics, Speech and Signal Processing (ICASSP)}. IEEE, 2023, pp. 1--5.

\bibitem{peng2023icassp}
Junyi Peng, Marc Delcroix, Tsubasa Ochiai, Old\v{r}ich Plchot, Shoko Araki, and Jan \v{C}ernocký,
\newblock ``Target speech extraction with pre-trained self-supervised learning models,''
\newblock in {\em ICASSP 2024-2024 IEEE International Conference on Acoustics, Speech and Signal Processing (ICASSP)}. IEEE, 2024.

\bibitem{delcroix2020improving}
Marc Delcroix, Tsubasa Ochiai, Katerina Zmolikova, Keisuke Kinoshita, Naohiro Tawara, Tomohiro Nakatani, and Shoko Araki,
\newblock ``{Improving Speaker Discrimination of Target Speech Extraction With Time-Domain Speakerbeam},''
\newblock in {\em ICASSP 2020-2020 IEEE International Conference on Acoustics, Speech and Signal Processing (ICASSP)}. IEEE, 2020, pp. 691--695.

\bibitem{radford2023robust}
Alec Radford, Jong~Wook Kim, Tao Xu, Greg Brockman, Christine McLeavey, and Ilya Sutskever,
\newblock ``Robust speech recognition via large-scale weak supervision,''
\newblock in {\em International Conference on Machine Learning}. PMLR, 2023, pp. 28492--28518.

\bibitem{luo2018tasnet}
Yi~Luo and Nima Mesgarani,
\newblock ``Tasnet: time-domain audio separation network for real-time, single-channel speech separation,''
\newblock in {\em 2018 IEEE International Conference on Acoustics, Speech and Signal Processing (ICASSP)}. IEEE, 2018, pp. 696--700.

\bibitem{snyder2018x}
David Snyder, Daniel Garcia-Romero, Gregory Sell, Daniel Povey, and Sanjeev Khudanpur,
\newblock ``{X-vectors: Robust DNN embeddings for speaker recognition},''
\newblock in {\em ICASSP 2018}. IEEE, 2018, pp. 5329--5333.

\bibitem{vzmolikova2019speakerbeam}
Kate{\v{r}}ina {\v{Z}}mol{\'\i}kov{\'a}, Marc Delcroix, Keisuke Kinoshita, Tsubasa Ochiai, Tomohiro Nakatani, Luk{\'a}{\v{s}} Burget, and Jan {\v{C}}ernock{\`y},
\newblock ``Speakerbeam: Speaker aware neural network for target speaker extraction in speech mixtures,''
\newblock {\em IEEE Journal of Selected Topics in Signal Processing}, vol. 13, no. 4, pp. 800--814, 2019.

\bibitem{cosentino2020librimix}
Joris Cosentino, Manuel Pariente, Samuele Cornell, Antoine Deleforge, and Emmanuel Vincent,
\newblock ``Librimix: An open-source dataset for generalizable speech separation,''
\newblock {\em arXiv preprint arXiv:2005.11262}, 2020.

\bibitem{nagrani2017voxceleb}
Arsha Nagrani, Joon~Son Chung, and Andrew Zisserman,
\newblock ``{VoxCeleb: A Large-Scale Speaker Identification Dataset},''
\newblock {\em Proc. Interspeech 2017}, pp. 2616--2620, 2017.

\bibitem{kolbaek2020loss}
Morten Kolb{\ae}k, Zheng-Hua Tan, S{\o}ren~Holdt Jensen, and Jesper Jensen,
\newblock ``On loss functions for supervised monaural time-domain speech enhancement,''
\newblock {\em IEEE/ACM Transactions on Audio, Speech, and Language Processing}, vol. 28, pp. 825--838, 2020.

\bibitem{delcroix2022listen}
Marc Delcroix, Keisuke Kinoshita, Tsubasa Ochiai, Katerina Zmolikova, Hiroshi Sato, and Tomohiro Nakatani,
\newblock ``{Listen only to me! How well can target speech extraction handle false alarms?},''
\newblock in {\em Proc. Interspeech 2022}, 2022, pp. 216--220.

\bibitem{kahn2020libri}
Jacob Kahn, Morgane Rivi{\`e}re, Weiyi Zheng, Evgeny Kharitonov, Qiantong Xu, Pierre-Emmanuel Mazar{\'e}, Julien Karadayi, Vitaliy Liptchinsky, Ronan Collobert, Christian Fuegen, et~al.,
\newblock ``Libri-light: A benchmark for asr with limited or no supervision,''
\newblock in {\em ICASSP 2020-2020 IEEE International Conference on Acoustics, Speech and Signal Processing (ICASSP)}. IEEE, 2020, pp. 7669--7673.

\end{thebibliography}

\end{document}